# Key Challenges for AAS Journals in the Next Decade[*]

Authors: AAS Publications Committee (Emily M. Levesque, Lisa Prato, Christopher Sneden, Jason W. Barnes, Dawn M. Gelino, Barbara Kern, Paula Szkody, Rosemary F. G. Wyse, Leslie A. Young)

**Summary**
- The American Astronomical Society (AAS) Journals are a vital asset of our professional society.
- With the push towards open access, page charges are a viable and sustainable option for continuing to effectively fund and publish the AAS Journals.
- The existing page charge model, which requires individual authors to pay page charges out of their grants or even out of pocket, is already challenging to some researchers and could be exacerbated in the Open Access (OA) era if charges increase.
- A discussion of alternative models for funding page charges and publishing costs should be part of the Astro2020 decadal survey if we wish to continue supporting the sustainable and accessible publication of US research in AAS journals in the rapidly-shifting publication landscape.
- The AAS Publications Committee recommends that the National Academy of Sciences form a task force to develop solutions and recommendations with respect to the urgent concerns and considerations highlighted in this White Paper.

## I. The Role of the AAS Journals

The American Astronomical Society journals - ApJ, AJ, ApJ Letters, and the ApJ Supplement Series - are a vital asset of our professional society and the most widely-read and frequently-cited journals in the field of astronomy[1]. The AAS is exceptional among professional scientific societies; we internally maintain and publish our own high-profile journals, a practice dating back to 1941. The Lead Editors and Scientific Editors of the journals are all drawn from the professional astronomy community, and independent oversight of the journals is carried out by the AAS Publications Committee, a group of 9-10 astronomers nominated and elected by committee members and approved by the AAS Board of Trustees. This leadership structure allows the journals to dynamically grow and change in response to community feedback and subject matter needs (for example, the inclusion of research on software and large databases in the "Instrumentation, Software, Laboratory Astrophysics, and Data" corridor) and continue serving the needs of our evolving field.

The AAS Journals maintain the valuable combination of both a high impact factor (5.55 for ApJ in 2017[2]) and a high acceptance rate (~85% for ApJ in 2017) for our journals. This publishing model supports the rigorous scientific standards upheld by the peer review process, making it possible to publish research that has withstood scrutiny by experts in the field, addresses important research questions, and contributes to the larger body of knowledge[3]. At the same time, the high acceptance rate and impact factor makes it possible for scientists at all career levels to have their research peer-reviewed, published, read, and cited.

---

[*] **The AAS Board of Trustees reviewed this White Paper and voted to endorse its content.**

The AAS Journals are also a valuable means of archiving and preserving research and data. The journals encourage authors to publish their data via large-scale machine readable tables and the "Data Behind the Figure" feature for sharing .FITS files and other material presented in publications[4]. The AAS Journals also encourage members of the community to publish articles on developing astronomical software, release open-source code, and cite software work where appropriate, making these journals an important source for researchers seeking to use cutting-edge software and other analysis techniques[5]. Grant requirements such as the NASA data management plans ask applicants to give concrete examples of how they will share the products of their research and preserve and disseminate results and data; the AAS Journals are a vital tool for accomplishing this and are regularly cited by astronomers as one of their primary means of satisfying these requirements.

**The continued publication of the AAS Journals is a substantial asset to the professional astronomy community and critical to its success.**

**II. Publishing Costs, Open Access, and the Role of Author Page Charges**
The costs associated with operating our own journals within AAS currently include direct and indirect support for editorial and production personnel, oversight e.g. Publications Committee costs, management, software systems for manuscript submission and processing, peer review, copyediting, special processing of digital content publication (including graphics, video, and software), online dissemination of content, archiving, curation and compliance costs, and more. The AAS journals are a non-profit, as opposed to large commercial publishers like Springer and Elsevier, the latter of which posted a 37.1% profit margin in 2018 (higher than tech giants such as Google and Amazon)[6,7].

As the AAS is not a commercial publisher, funding to support the journals' operating costs must come from a combination of subscription fees and page charges. **No AAS membership dues, meeting fees, or other sources of revenue are allocated for publishing costs.** An annual institutional subscription to all four of the AAS Journals (typically purchased by college and university libraries for use by all members and affiliates of the institution) currently costs $2,649. Page charges for the AAS journals are currently parsed into "quanta" and charged at $30 per 350 words + $35 per digital quantum (e.g. a figure, table, or other digital data component); the median charge for a AAS Journals article in 2018 was $928. These charges are currently paid by individual authors as a cost of publishing their research in the AAS journals. At the time of manuscript submission, authors may request a waiver of page charges in special circumstances. The Editor in Chief will consider these requests on a case by case basis after completion of the scientific review process.

Page charges in particular are cited as a problem by authors and a reason to publish elsewhere. In a recent survey of 3,663 authors of astronomy papers, 55% of them cited the lack of page charges as highly influential in their decision to publish work in journals with no author charges (e.g. MNRAS, Nature Astronomy, Icarus), and this number increased to 64% when surveying only early-career authors[8]. However, page charges make up ~65% of the AAS Journals' revenue, with subscriptions making up the remaining ~35%. Moving to a zero author charge model for the AAS Journals would raise the cost of a yearly institutional subscription to at least $7,950[7].

There has also been a push in recent years to shift the academic publishing model to what is known as Open Access (OA), a structure in which publications are freely available to the public without any subscription or per-article charges. This follows broad calls for increasing the accessibility of scientific data and results and a dedicated push in Europe towards exclusively OA publishing. Science Europe (a consortium of EU national science funding agencies) recently announced "Plan S", which demands that all scholarly publications supported by public funding be freely available to the public by 2024. OA is widely seen as a positive step in publishing, removing the access barrier of page charges and allowing the public access to research funded by their tax dollars. However, because this model removes subscriptions as a revenue stream, OA typically results in an increase in article processing charges or page charges[9].

The AAS journals currently operate on what is known as a "hybrid" OA model; the default "Green OA" model for journal publications makes them accessible to the public without a subscription after 12 months. Authors also can choose to publish their paper as "Gold OA" for an additional fee that offsets the lost subscription revenue, and the AAS journals' copyright permits authors to also post article preprints online to make them available to the public at no cost. However, this hybrid OA model does not comply with Plan S; under the current provisions, authors would not be able to publish work supported by EU funding agencies in the AAS Journals.

Some journals are able to avoid page charges for their authors, but this typically comes at a cost. Large for-profit publishers such as Springer and Elsevier have much higher subscription prices than the AAS Journals, a model that passes the bulk of publishing expenses to institutional libraries. This in turn disproportionately impacts public institutions and smaller colleges and universities, whose libraries (often amidst decreasing budgets) must choose between paying an increasingly expensive (and potentially unmanageable) subscription fee or not subscribing and instead leaving researchers at their institutions to pay per article or rely on their libraries' interlibrary loan agreements for access to published research.

The most commonly-cited example of a journal without page charges in astronomy is the Monthly Notices of the Royal Astronomical Society (MNRAS). However, the current electronic-only subscription price for MNRAS in the United States is $10,249, and the journal is part of the expensive Oxford University Press bundle. Under the increasing financial strain of academic publishing, even MNRAS has recently updated their page charge policy: in August 2018 they began charging authors for papers over 20 pages at a rate of £50 (about $63) per additional page.

In short, page charges currently remain the most viable model for financially supporting academic publishing, especially as OA publishing grows in popularity.

**III. The Challenges of Author Page Charges**

Individuals typically pay for their page charges using grant funding, and for US authors the majority of this funding in astronomy comes from federal organizations such as the NSF and NASA. Success rates for grant applications have decreased in astronomy (the success rate in the NSF's AAG program has dropped from 50.4% in 1990 to 20.9% in 2016, and was as low as 14.8% in 2012[10]), making it increasingly difficult for individual investigators to pay page charges. The total cost of page charges to an individual research grant is also higher when considering that most publicly-funded grants are charged overheads by their host institutions; these overhead rates now typically exceed 50% at public institutions and can be even higher for private colleges and universities. To quantify this, in 2018 an author at an institution with an overhead rate of 55.5% who wished to publish a paper in an AAS journal with Gold OA would have had to budget a median of $3,337.

The current page charge model is particularly burdensome for authors at smaller institutions, including small liberal arts colleges, community colleges and minority-serving institutions, as grant overheads may be higher and institutional funds aren't necessarily widely available. Page charges will also disproportionately impact women and underrepresented minorities, who experience well-documented discrimination in the grant and proposal review process and will have lower funding rates as a result (a 2014 study found a systemic bias in favor of Hubble proposals submitted by male PIs[11]). The same impact is felt by authors from outside the US who want to publish in the AAS journals, particularly those coming from underrepresented countries where research funding is limited. As a whole, our existing system of paying page charges perpetuates existing inequities in scientific research by making it more difficult for authors from underrepresented groups and institutions to publish their science.

In an effort to find an affordable publishing option or to limit grant expenses, these pressures often drive individual authors to publish their work elsewhere rather than within the AAS journals. This may include publishing in cheaper but lower-impact journals (which could have negative consequences for authors seeking jobs, promotion, and/or tenure), or publishing in non-OA/for-profit journals with no page charges, an option that will diminish as OA publishing becomes more prevalent. The result is a system that will, in the long term, negatively impact both the AAS journals and the US astronomy community.

### IV. Astronomical Publishing in the 2020's

The current AAS Journals publishing model has allowed us to maintain an excellent scientific resource in the astronomy community. The journals have both a high acceptance rate and a high impact factor and are a widely-used and valuable means of disseminating research results and resources throughout the community. However, the current model is also dependent on author page charges. This approach is already challenging to some members of the community and will become even more difficult to sustain in the 2020's as the push toward OA continues, a move that will ultimately eliminate subscriptions as a source of revenue. Left unchanged, the current AAS Journals' funding model would then depend entirely on author page charges, which would need to be raised significantly to continue offsetting publication costs.

Authors, AAS members, and the broader community need to address these and other concerns as part of a larger plan for both the near- and long-term future of astronomical publishing. Page charges are and will remain an important source of revenue for the AAS Journals, a fact that will only become more true as the move towards gold OA continues. However, we believe that as part of the decadal survey we must consider options for funding page charges that can lessen the financial burden on individual U.S. authors' budgets and support OA publishing in a manner that continues to support the mission of the AAS Journals and their continued success in a dynamic international publishing environment.

**References**
 *1 -* AAS Publishing journal summary (*link*)
 *2 -* IOP Publishing (*link*)
 *3 -* Kelly et al. 2014, EJIFCC, PMID 27683470 (*link*)
 *4 -* AAS Publishing Data Guide (*link*)
 *5 -* AAS Publishing Policy Statement on Software (*link*)
 *6 -* Page, B. "Elsevier records 2% lifts in revenue and profits" thebookseller.com (*link*)
 *7 -* AAS Publishing Business Model (*link*)
 *8 -* AAS Publications Committee author survey
 *9 -* Laakso et al. 2016, Learned Published, 29, 259 (*link*)
 *10 -* Green, R. AAS 231 NSF/AST Town Hall (*link*)
 *11 -* Reid, N. "Gender-based Systematics in HST Proposal Selection (*link*)